%% file: TGF_macro.tex
\documentclass{svmult}

\usepackage{makeidx}         
\usepackage{graphicx}        
\usepackage{multicol}        
\usepackage[bottom]{footmisc}

\input{defsSubmit}

\makeindex             

\begin{document}

\title*{Crowd Flow Modeling of Athletes in Mass Sports Events
- a Macroscopic Approach}
\titlerunning{Macroscopic Crowd Flow Modeling of Athletes}

\author{Martin Treiber\inst{1}}

\institute{Technische Universit\"at Dresden, 
D-01062 Dresden
\texttt{treiber@vwi.tu-dresden.de}
}
%
%

\maketitle



\begin{abstract}
.
We propose a macroscopic model in form of a 
dispersion-transport equation for non-congested flow of the athletes which is coupled to 
a kinematic-wave model for congested flow. The model takes into
account the  performance (i.e., free-flow
speed distributions) of the athletes in the different starting
groups. The model is calibrated and validated on 
 data of the German
\emph{Rennsteig Half Marathon 2012} and the Swedish \emph{Vasaloppet
  2012} cross-country ski race. 
Simulations of the model allow the event managers to improve the
organization by determining the optimum number of starting groups,
the maximum size of each
group, whether a wave start with a certain starting delay between
the groups is necessary, or what will be the effects of changing the
course. We apply the model to simulate a planned course change for the
Rennsteig Half Marathon 2013, and determine whether critical
congestions  are likely to occur.
\end{abstract}

\section{Introduction}
\begin{figure}
\fig{\textwidth}{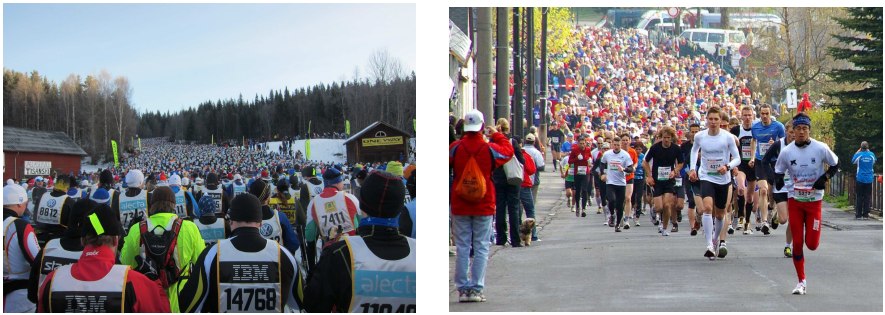}
\caption{\label{fig:photos}Jams at the Swedish Vasaloppet
  cross-country race 2012 (left) and at the Rennsteig Marathon 2012
  (right).
}
\end{figure}

%
Mass-sport events for runners, cross-country skiers, or other athletes,
are increasingly popular. Prominent examples include the New York Marathon, the Vasaloppet
cross-country ski race in Sweden, and the nightly inline-skating events taking place
in nearly every major European city. Due to their popularity (the number
of participants is typically in the thousands, sometimes in the ten thousands),
``traffic jams'' occur regularly (Fig.~\ref{fig:photos}). They are not
only a hassle for the 
athletes (since the time is ticking) but also pose organisational or
even safety threats, e.g., because a spillback from a jam threatens to
overload a critical bridge. Nevertheless, scientific investigations of the
athletes' crowd flow 
dynamics~\cite{TreiberKesting-Book} are virtually nonexisting. 

The crowd dynamics can be described by two-dimensional
active-particle systems~\cite{Helbing-01aa}. Unlike the situation in
general pedestrian 
traffic, the flow is unidirectional since all athletes share the same
destination (the finishing line). This means, the
dynamics is equivalent to that of mixed unidirectional
vehicular traffic flow
 which may be lane-based,
as in cross-country ski races in the classic style~\cite{TGF13-ski},
or not, as in running events 
but also in mixed vehicular traffic flow in many developing
countries~\cite{Arasan-mixedTraffic}. 
The uni-directionality allows to simplify the mathematical
description to a macroscopic, one-dimensional model for the motion
along the  longitudinal (arc-length)
coordinate.

In this contribution, we formulate  a macroscopic
dispersion-transport model for free flow which is coupled to a
kinematic-wave model for congested flow. We calibrate and validate
the model by data of the Rennsteig 2012 Half Marathon
and the Vasaloppet 2012 and apply it to
simulate the effects of a planned course change for the next
Rennsteig Half Marathon 2013 to avoid the overloading of a critical bridge. 

In the next section, we develop the macroscopic model and 
show its workings on data of past running and ski
events. In Section~\ref{sec:appl}, we apply it to simulate
organisational changes for the Rennsteig Half
Marathon 2013. Finally,
Sec.~\ref{sec:discuss} gives a discussion.

\section{\label{sec:model}The Macroscopic Model}
%
Our proposed macroscopic model has two components for free and
congested traffic, respectively. Since, 
in free traffic, individual performance differences
translate into different speeds, we formulate the free-traffic part as
a multi-class model. In contrast, ``everybody is treated equal''
in congested traffic, so a simple single-class kinematic-wave 
model is sufficient. During the simulation, the free-traffic part provides
the spatio-temporally changing traffic demand (athletes per second). A congestion arises
as soon as the local demand exceeds the local capacity. The
resulting moving upstream boundary of the jam is subsequently described by
standard shock-wave kinematics.

\subsection{Free Traffic Flow}
\begin{figure}
\fig{\textwidth}{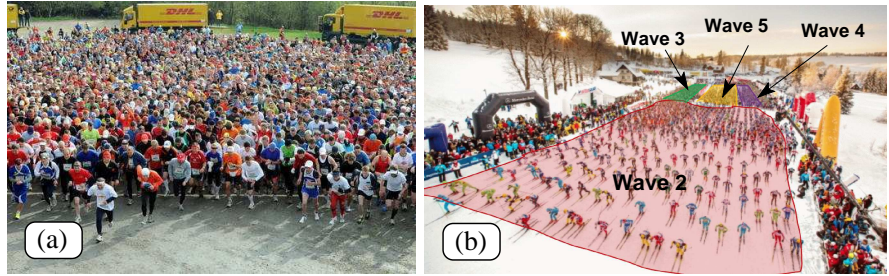}
\caption{\label{fig:start}Two possible starting schemes. (a) mass
  start (Rennsteig Marathon 2012); (b) wave start (Jizerska
  Padesatka \unit[50]{km}, 2012). 
}
\end{figure}

In most bigger mass sports events, the athletes are classified
according to performance into
starting groups.  All groups start  either 
simultaneously (``mass start'', Fig.~\ref{fig:start} (a)), or
sequentially with fixed delays between the groups which, then,
are also called waves (``wave start'', Fig.~\ref{fig:start} (b)).  

Generally, each athlete wears an individual RFID chip recording the
starting and finishing time, and also split times when passing
refreshment stations along the course.
The information of the starting groups is highly useful since the
speed distribution within
each group is much narrower than that for the complete field. Thus,
by considering each group individually, the model makes
more precise predictions. 
\begin{figure}
\fig{\textwidth}{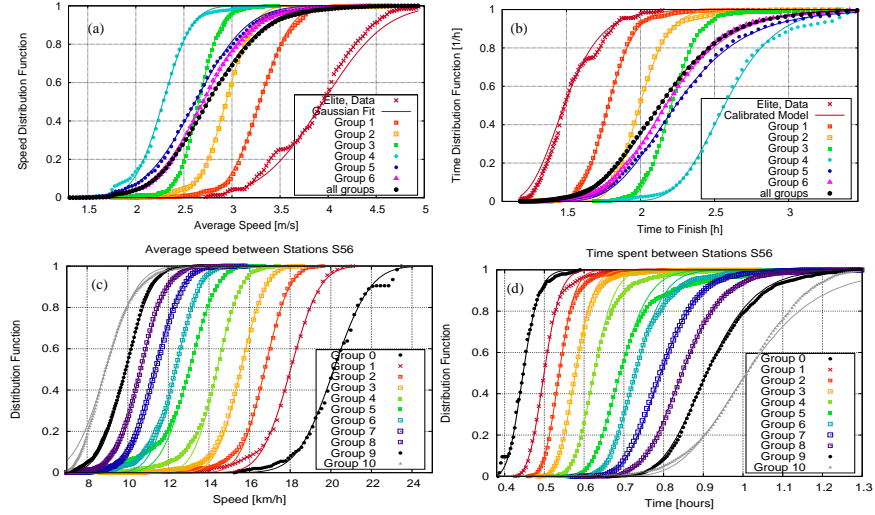}
\caption{\label{fig:distr}Distribution of the average speeds (left)
  and times (right) of the different starting groups at the  Rennsteig Half
  Marathon 2012 (whole race, top) and at the Vasaloppet~2012
  (a \unit[9]{km} section between the
  stations~5 and~6, bottom). Symbols: data; curves: model.
}
\end{figure}

Figure~\ref{fig:distr} shows the
distributions of the final times of the German
Rennsteig Half Marathon 
and the time for a section of the Vasaloppet 2012 where no major jams
are observed. We fitted the data of each group by Gaussians
parameterized, for reasons of robustness,
 by the median and the inter-quartile gap instead of the
arithemic mean
and standard deviation. We infer that, in the absence of major
disturbances, the speed distribution
within each group is nearly Gaussian. 
Significant deviations are only observed (i) for the small
elite groups due to platooning, (ii) for the low-speed
tails. (Generally, the low-speed tails are fatter compared to
Gaussians. However, at the
Vasaloppet, the slowest athletes are taken out of the race thus reversing
this effect.)

Using the normal kinematic relation $T=L/v$ for the time $T$
that athletes of group $k$ take to cover the 
distance $L$ at speed $v$, we obtain  by
elementary probability theory following relation between 
the density functions $f^v_k(v)$ of the speed and the (non-Gaussian) density function
$f^T_k(T|L)$ of the needed time,
\be
\label{Tdistr}
 f^T_k(T|L)=\frac{L}{T^2} f^v_k\left(\frac{L}{T}\right).
\ee

\begin{figure}
\fig{0.75\textwidth}{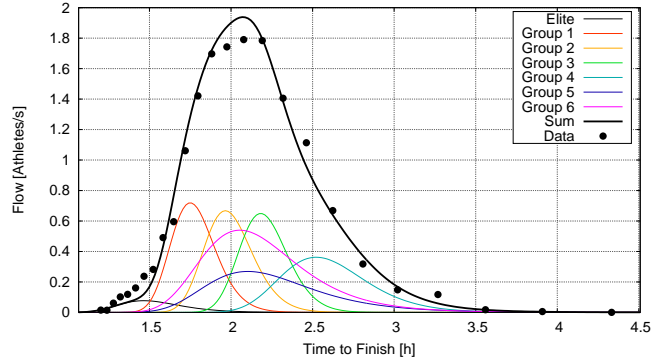}
\caption{\label{fig:partialFlows} Partial flows $Q_k(x,t)$ of the different
  starting groups (mass start) at the finish at \unit[21.1]{km} (colored curves)
  and total flow according to~\refkl{QfreeWave} (thick curve) compared with
  the actual flow (data points). 
}
\end{figure}

Finally, we assume that the relative performance of an athlete
persists throughout the race. In other words, in free traffic, 
a fast runner remains
fast and a slow athlete slow. This means, the flow dynamics obeys a
dipersion rather than a diffusion equation. Specifically, we assume constant
speed distributions on flat terrain and identical relative speed changes for
inhomogeneities such as uphill or downhill gradients. In the
following, we will assume a flat terrain, for notational simplicity.

\begin{figure}
\fig{\textwidth}{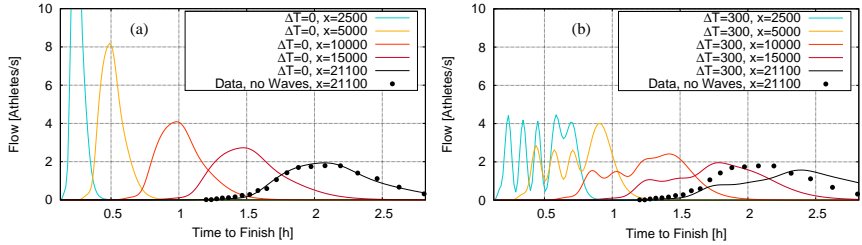}
\caption{\label{fig:waveStart}Simulated effect of a wave start on
  the local flow of athletes (all starting groups) at
  different locations from $x=\unit[2\,500]{m}$ to the finish. (a)
  Reference (mass start), (b) delay of 
  \unit[5]{min}.
}
\end{figure}

Denoting the number of athletes in each 
group by $n_k$
and assuming a wave start where group $k$ 
 starts a time delay $\tau_k$ after the starting gun goes off
 (indicating the start of the first and elite
waves), the free-traffic demands $Q\sub{free}(x,t)$ and
densities $\rho\sub{free}(x,t)$ read
\bea
\label{QfreeWave}
Q\sub{free}(x,t) &=& \sum_{k} Q_k(x,t-\tau_k), 
\quad Q_k(x,t)= n_k f^T_k(t|x), \\
\label{rhofreeWave}
\rho\sub{free}(x,t) &=& \sum_{k} \rho_k(x,t-\tau_k),
\quad \rho_k(x,t)=\frac{t}{x} \, Q_k(x,t),
\eea
where $Q_k$ and $\rho_k$ are set to zero for time arguments
$t-\tau_k\le 0$. Figure~\ref{fig:partialFlows} shows that the model
prediction for the total traffic demand $Q\sub{free}(x,t)$ at the
finish line fits well 
with the data (possibly, the small deviation at the peak is due to
congestions). Thus, we are now able to estimate the free-flow traffic
demand upstream of a congestion at
any location and at any time during the race. Moreover, we now can
anticipate the consequences of organisational changes such as
realizing a wave
start rather than a mass start (Fig.~\ref{fig:waveStart}).

\subsection{Kinematic-Wave Model for Congested Crowds}
We propose a quasi-onedimensional Lighthill-Whitham-Richards model
with a triangular fundamental diagram. In terms of the local capacity $C(x)$
(maximum number of athletes per second that can pass a cross section
at location $x$),
the free-flow speed $V_0$, and the maximum local one-dimensional density
(athletes per meter) $\rho\sub{max}(x)$, the fundamental diagram can be
expressed by~\cite{TreiberKesting-Book}
\be
\label{funddia}
Q_e(\rho)=\max\left[V_0\rho, 
 \frac{C (\rho\sub{max}-\rho)}{\rho\sub{max}-C/V_0}\right].
\ee
Notice that the observed capacity $C$ increases weakly with the maximum
speed $V_0$ such that $V_0$ essentially cancels out in the congested
branch of~\refkl{funddia}. 
A traffic breakdown arises if, at any location or time, the free-flow
demand $Q\sub{free}(x,t)$ exceeds the local
capacity $C_B$ at a bottleneck $x=x_B$ (where the capacity is at
a local minimum). The resulting congested
traffic region has a one-dimensional density
\be
\label{rhoc}
\rho\sub{cong}(x)=\rho\sub{max}(x)\left(1-\frac{C_B}{C(x)}\right)+\frac{C_B}{V_0}.
\ee
The congestion has a
 stationary downstream front at the bottleneck
location $x_B$ while the upstream front $x\sub{up}(t)$ is moving according to the
shock-wave formula
\be
\label{shock}
\abl{x\sub{up}}{t}=\frac{C_B-Q\sub{free}(x,t)}{\rho\sub{cong}(x)-\rho\sub{free}(x,t)}.
\ee
The congestion dissolves as soon as $x\sub{up}(t)$ crosses $x_B$ in
the downstream direction. Finally, the free-traffic flow downstream of the
congested region has a constant flow $Q\sub{free}\sup{down}(x,t)=C_B$ equal to
the bottleneck capacity.

Both the local capacities and maximum densities are
proportional to the local width $w(x)$ of the course:
\be
\label{C}
C(x)=J\sub{max} w(x), \quad
\rho\sub{max}(x)=\rho\sub{max}\sup{2d}w(x).
\ee
The maximum flow density (specific capacity) $J\sub{max}$
and the maximum 2d density $\rho\sub{max}\sup{2d}$ are model
parameters depending on the kind of race and on the local conditions
(e.g., gradients). From past congestions, we can estimate
$\rho\sub{max}\sup{2d}=\unit[2]{m^{-2}}$ and
$J\sub{max}=\unit[1.5]{(ms)^{-1}}$ for running competions on level
terrain (which is comparable to normal unidirectional
pedestrian flows),  and
$\rho\sub{max}\sup{2d}=\unit[0.7]{m^{-2}}$, 
$J\sub{max}=\unit[0.6]{(ms)^{-1}}$ for level-terrain cross-country
ski events.

\section{\label{sec:appl}Simulating Scenarios for a
  Marathon Event}
%
At the 2012 Rennsteig Half Marathon, there were six starting groups.
The last group contained significantly more participants. For
2013, the managers plan eight groups of equal size $n_k\le 850$, with
the first five groups sorted to performance, and the last three groups
available for
the runners for which no previous performance are known or who
registered too late. Based on the
2012 data, we set the average
speeds to  $v_1=\unit[3.5]{m/s}$, 
$v_2=\unit[3.1]{m/s}$, $v_3=\unit[2.7]{m/s}$, $v_4=\unit[2.4]{m/s}$,
$v_5=\unit[2.1]{m/s}$, and $v_6=v_7=v_8= \unit[2.7]{m/s}$. All speed
variances are assumed to be $\sigma_v^2=\unit[0.15]{m^2/s^2}$. 

\begin{figure}
\fig{0.7\textwidth}{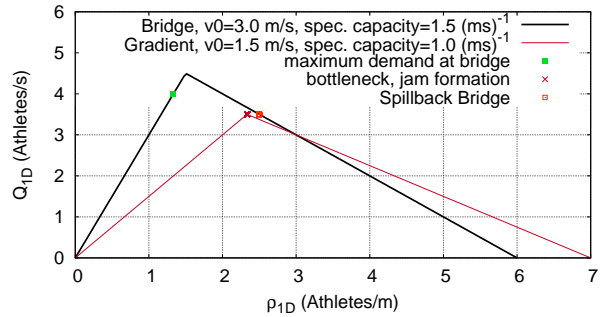}
\caption{\label{fig:funddia}Fundamental diagram for different
  situations of the simulation of the Rennsteig Half Marathon 2013 (see
  the main text for details.)
}
\end{figure}

Due to external constraints, the course of the 2013 Marathon must be
changed. There are several options:
\bi
\item Scenario~1: Mass start. The \unit[5]{m} wide starting section has a capacity of
  7~athletes/s. The first bottleneck at $x=\unit[1\,000]{m}$ is a
  7\% uphill gradient section of \unit[4.5]{m} width. At
  $x=\unit[2\,200]{m}$, the athletes encounter a \unit[3.5]{m} wide
  downhill section. The critical bottlenecks, however, consist of a
  bridge at $x=\unit[3\,000]{m}$ (level, \unit[3]{m} wide), and,
  \unit[100]{m} afterwards, a
  steep uphill gradient (11\%) where the course has a width of
  \unit[3.5]{m}. 
\item Scenario~1a: As Scenario~1, but wave start with a delay of
  \unit[300]{s} per wave
\item Scenario~1b: As Scenario~1a, but the capacity of the starting
  section has been reduced to 5.5~athletes/s.
\item Scenario~2: The course is reorganized such that the 7\% gradient
  is at $x=\unit[1\,400]{m}$, the downhill bottleneck at
  $x=\unit[2\,700]{m}$, and the bridge with the subsequent steep uphill
  section at $x=\unit[5\,700]{m}$ and \unit[5\,800]{m}, respectively.
\ei
Based on past experience, the maximum 2d density is set to
$\rho\sub{max}\sup{2d}=\unit[2]{m^{-2}}$ and the specific capacities to
$J\sub{max}=\unit[1.5]{(ms)^{-1}}$ for level sections
(including the bridge), and $\unit[1.2]{(ms)^{-1}}$,
$\unit[1.0]{(ms)^{-1}}$, and $\unit[1.3]{(ms)^{-1}}$
for the 7\%, 11\%, and the downhill gradients, respectively. 
Figure~\ref{fig:funddia} displays the resulting fundamental diagrams
for the bridge (capacity $C=\unit[4.5]{s^{-1}}$) and the subsequent
uphill section ($C=\unit[3.5]{s^{-1}}$)

While some congestions are unavoidable, we must require that
there is no significant congestion on the \unit[60]{m} long bridge
itself because this may result in dangerous overloading.

\begin{figure}
\fig{\textwidth}{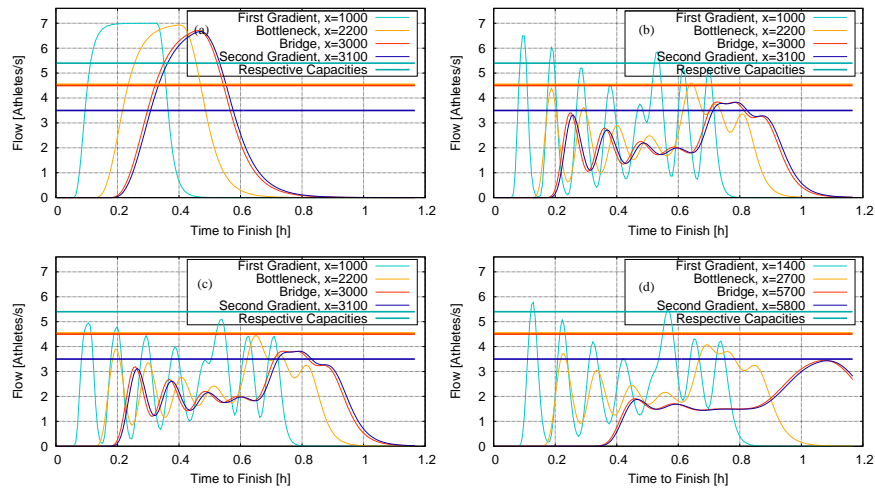}
\caption{\label{fig:scenarios}Two simulated scenarios for the
  Rennsteiglauf Half Marathon 2013. The horizontal lines give the
  capacities of the various bottlenecks of the course, and the curves
  of the same color the predicted demands at these positions.
}
\end{figure}

Figure~\ref{fig:scenarios} shows the main results: With a mass start
(Fig.~\ref{fig:scenarios}(a)), massive jams will form at and upstream of all
the bottlenecks, including a spillback to the bridge, so this is no
option. Adopting a wave start (Fig.~\ref{fig:scenarios}(b)) reduces
the congestion at the first bottleneck to a
tolerable level. Furthermore, jams are no longer expected at the downhill
bottleneck while the bridge itself has even capacity to
spare. However, the demand exceeds the capacity of the
steep uphill section leading to a supply-demand mismatch of up to
about 150 athletes (the area 
between the blue curve and the blue capacity line of
Fig.~\ref{fig:scenarios}(b)). This corresponds to a jam of about
\unit[150]{m}, i.e., there is a spillback with a density of
\unit[2.5]{athletes/m} (cf. Fig.~\ref{fig:funddia}) onto the bridge. 
Reducing the  initial capacity of
the starting field to 5.5 athletes/s (Fig.~\ref{fig:scenarios}(c))
does not help much in this situation. Only a rearrangement of the
course with the bridge section located further away from the start
yields a significant improvement with the only (minor) jam expected at the
uncritical first bottleneck.

\section{\label{sec:discuss}Discussion}
%
We have proposed a  macroscopic dispersion-transport model that allows managers of
mass-sports events to assess the implications of changing
the course, or the spatio-temporal organization of the start, without
prior experiments.  
As a general rule, critical bottlenecks should be moved as far away
from the start as possible. If the situation remains critical, a wave
start and/or a restriction of the number of participants will be
necessary.



\input{TGF_macro.bbl}
\end{document}

%% file: defsSubmit.tex
\usepackage{units} 

\providecommand{\bc}{\begin{center}}
\providecommand{\ec}{\end{center}}
\providecommand{\be}{\begin{equation}}
\providecommand{\ee}{\end{equation}}
\providecommand{\bea}{\begin{eqnarray}}
\providecommand{\eea}{\end{eqnarray}}
\providecommand{\bdm}{\begin{displaymath}}
\providecommand{\edm}{\end{displaymath}}
\providecommand{\bdma}{\begin{eqnarray*}}
\providecommand{\edma}{\end{eqnarray*}}
\providecommand{\ba}{\begin{eqnarray*}}
\providecommand{\ea}{\end{eqnarray*}}
\providecommand{\bi}{\begin{itemize}}
\providecommand{\ei}{\end{itemize}}
\providecommand{\benum}{\begin{enumerate}}
\providecommand{\eenum}{\end{enumerate}}

\providecommand{\refkl}[1]{(\ref{#1})}

\providecommand{\text}[1]{{\mbox{ #1}}}

\providecommand{\fig}[2]{
   \begin{center}
     \includegraphics[width=#1]{#2}
   \end{center}
}

\providecommand{\abl}[2]{\frac{{\rm d} #1}{{\rm d} #2}}  

\providecommand{\sub}[1]{_{\rm #1}}
\renewcommand{\sup}[1]{^{\rm #1}}

